\documentclass[twocolumn,preprintnumbers,amsmath,amssymb,prl, superscriptaddress]{revtex4}
\pdfoutput=1
\usepackage{latexsym}
\usepackage{amssymb}
\usepackage{epsfig,amsmath,graphics}
\usepackage{epstopdf}
\usepackage{verbatim}
\usepackage[hypertex]{hyperref}
\usepackage{graphicx}
\usepackage{subfigure}

\newcommand{\GeV}{\text{GeV}}
\newcommand{\TeV}{\text{TeV}}

\newcommand{\s}{\text{s}}

\newcommand{\cm}{\text{cm}}
\newcommand{\sr}{\text{sr}}
\newcommand{\kpc}{\text{kpc}}
\newcommand\ee{\end{equation}}
\newcommand\be{\begin{equation}}

\newcommand{\OrderOne}{\mathcal{O}\left(1\right)}

\newcommand{\Mgut}{M_\text{GUT}}
\newcommand{\gut}{\text{GUT}}

\def\r{\right)}
\def\l{\left(}

\begin{document}


\title{Decaying Dark Matter as a Probe of Unification and TeV Spectroscopy}

\author{Asimina Arvanitaki}
\affiliation{Berkeley Center for Theoretical Physics, University of California, Berkeley, CA, 94720}
\affiliation{Theoretical Physics Group, Lawrence Berkeley National Laboratory, Berkeley, CA, 94720}

\author{Savas Dimopoulos}
\affiliation{Department of Physics, Stanford University, Stanford, California 94305}

\author{Sergei Dubovsky}
\affiliation{Department of Physics, Stanford University, Stanford, California 94305}
\affiliation{ Institute for Nuclear Research of the Russian Academy of Sciences,
        60th October Anniversary Prospect, 7a, 117312 Moscow, Russia}

\author{Peter W. Graham}
\affiliation{Department of Physics, Stanford University, Stanford, California 94305}

\author{Roni Harnik}
\affiliation{Department of Physics, Stanford University, Stanford, California 94305}

\author{Surjeet Rajendran}
\affiliation{SLAC National Accelerator Laboratory, Stanford University, Menlo Park, California 94025}
\affiliation{Department of Physics, Stanford University, Stanford, California 94305}

\date{\today}

\begin{abstract}
In supersymmetric unified theories the dark matter particle can decay, just like the proton, through grand unified interactions with a lifetime of order of $\sim 10^{26} $ sec.  Its decay products can be detected by several experiments -- including Fermi, HESS,  PAMELA, ATIC, and IceCube -- opening our first direct window to physics at the TeV scale and simultaneously at the unification scale $\sim10^{16}$ GeV.  We consider possibilities for explaining the electron/positron spectra observed by HESS, PAMELA, and ATIC, and the resulting predictions for the gamma-ray, electron/positron, and neutrino spectra as will be measured, for example, by Fermi and IceCube.  The discovery of an isotropic, hard gamma ray spectral feature at Fermi would be strong evidence for dark matter and would disfavor astrophysical sources such as pulsars.  Substructure in the cosmic ray spectra probes the spectroscopy of new TeV-mass particles. For example, a preponderance of electrons in the final state can result from the lightness of selectrons relative to squarks. Decaying dark matter acts as a sparticle injector with an energy reach potentially higher than the LHC. The resulting cosmic ray flux depends only on the values of the weak and unification scales.
\end{abstract}

\maketitle

\section{Introduction}
In grand unified theories the proton can decay because the global baryon-number symmetry of the low energy Standard Model is broken by physics at the grand unification (GUT) scale.  Indeed, only local symmetries can guarantee that a particle remains exactly stable, whereas global symmetries are generally broken in fundamental theories.  Just as the proton is long-lived but may ultimately decay, other particles, for example the dark matter, may decay with long lifetimes. It would seem miraculous that the dark matter particle's lifetime is in a range which is long enough to be a good dark matter candidate yet short enough to be observable today.  Nevertheless, this may well be what happens in Grand Unified Theories.  For example if a TeV mass dark matter (DM) particle decays via GUT-suppressed dimension 6 operators, its lifetime would be
\begin{equation}
\label{Eqn: dim 6 lifetime}
\tau \sim 8 \pi \frac{\Mgut^4}{m_{\text{DM}}^5} = 3 \times 10^{27} ~\s \left( \frac{\TeV}{m_{\text{DM}}} \right)^5 \left( \frac{\Mgut}{2 \times 10^{16} ~\GeV} \right)^4
\end{equation}
where $\Mgut \approx 2 \times 10^{16} \, \GeV$ is the supersymmetric (SUSY) unification scale.

This lifetime is being probed by several current experiments, as shown in Table \ref{Tab: astro limits}.
This can be understood, at least for the satellite and balloon experiments, because these all generally have similar acceptances of $\sim (1 ~\text{m}^2) (1~ \text{yr}) (1 ~\sr) \approx 3 \times 10^{11} ~ \cm^2 ~\s ~ \sr$.  For comparison, the number of incident particles from decaying dark matter with a lifetime of $10^{27}$ s is $\sim \int^{10 ~\kpc} \frac{d^3r}{r^2}  (0.3 ~\frac{\GeV}{m_{\text{DM}} ~\cm^3}) (10^{-27} ~\s^{-1}) \approx 10^{-9} ~\cm^{-2} ~\s^{-1} ~\sr^{-1}$, where these could be photons, positrons or antiprotons for example, depending on what is produced in the decay.  This implies such experiments observe $\sim \left( 3 \times 10^{11} ~ \cm^2 ~\s ~ \sr \right) \times \left( 10^{-9} ~\cm^{-2} ~\s^{-1} ~\sr^{-1} \right) \approx 300$ events. This coincidence may allow these experiments to probe physics at the GUT scale, much as the decay of the proton and a study of its branching ratios would.  In fact, HESS, PAMELA, and ATIC may have preliminary indications of dark matter from the cosmic ray electron/positron spectrum.  The Fermi satellite will test this by measuring both the electron/positron and gamma-ray spectra with significantly improved precision.



\begin{table}
\begin{center}
\begin{math}
\begin{footnotesize}
\begin{array}{|c|c|c|}
\hline
\text{Decay Channel} & \tau \text{ Lower Limit} & \text{Experiment} \\
\hline
q \overline{q} & 10^{27} ~\s & \text{PAMELA antiprotons} \\
\hline
e^+ e^- \text{ or } \mu^+ \mu^- & 2 \times 10^{25} ~\s \left( \frac{\TeV}{m_{\text{DM}}} \right) & \text{PAMELA positrons} \\
\hline
\tau^+ \tau^- & 10^{25} ~\s \left(1 + \frac{\TeV}{m_{\text{DM}}} \right) & \text{EGRET + PAMELA} \\
\hline
W W & 3 \times 10^{26} ~\s  & \text{PAMELA antiprotons} \\
\hline
\gamma \gamma & 2 \times 10^{25} ~\s & \text{PAMELA antiprotons} \\
\hline
\nu \overline{\nu} & 10^{25} ~\s ~ \left( \frac{m_{\text{DM}}}{\TeV} \right) & \text{AMANDA, Super-K} \\
\hline
\end{array}
\end{footnotesize}
\end{math}
\caption[Astrophysical Limits on Decaying Dark Matter]{\label{Tab: astro limits} A lower limit on the lifetime of a dark matter particle with mass in the range $100 ~\GeV \lesssim m_{\text{DM}} \lesssim 10 ~\TeV$, decaying to the products listed in the left column.  The experiment and the observed particle being used to set the limit are listed in the right column.  All the limits are only approximate.  Generally conservative assumptions were made and there are many details and caveats as described in \cite{Arvanitaki:2008hq}.}
\end{center}
\end{table}


Previously \cite{Arvanitaki:2008hq}, we discussed the general framework of dark matter decays induced by GUT scale physics and its signals at PAMELA, ATIC and Fermi. In this paper, we consider the implications of the recently published HESS data \cite{HESS} on this framework and discuss models that fit the overall electron/positron spectrum observed by PAMELA, ATIC and HESS. We focus on the correlated photon and neutrino signals that could be observed at Fermi and IceCube respectively.  

\section{Theoretical setup}
To study the observational consequences of decaying dark matter in SUSY GUTs one may follow an effective field theory approach and consider
an extended MSSM with higher dimensional operators parametrizing GUT effects and leading to dark matter decay. A detailed analysis of possible higher dimensional operators and the ways to generate them from  concrete microscopic SUSY GUTs was presented in
 \cite{Arvanitaki:2008hq}. Here, for definiteness, we will work in the context of the $SO(10)$ models described in  \cite{Arvanitaki:2008hq}. 
As an example, in addition to the standard MSSM interactions, we introduce
an additional
vectorlike  $\l16_m, \bar{16}_m\r$ multiplet at the TeV scale and $10_{\gut}$ multiplet at the GUT scale.
The relevant superpotential interactions involving these fields are
\begin{equation}
\label{SO10GUT}
W^{\prime} = \lambda 16_m 16_f 10_{\gut} + m 16_m \bar{16}_m + \Mgut 10_{\gut} 10_{\gut}
\end{equation}
 We will assume that the singlet $S_m$ is the lightest component of the $16_m$ and will therefore be dark matter. After GUT scale matter and gauge fields are integrated out one obtains the dimension 5 
operator $16_m 16_m 16_f 16_f$ in the superpotential and dimension 5 and 6 Kahler terms
$16_m 16_m 10^{\dagger}$, $16^{\dagger}_m 16_m 16^{\dagger}_f 16_f$ involving $m$-fields. Of all these, the only operator that 
involves two singlet $S_m$ components of $16_m$ is the dimension 6 Kahler term yielding $S^{\dagger}_m S_m 16^{\dagger}_f 16_f$ (assuming right-handed neutrinos are heavy).
Consequently, in this model a thermal relic abundance of singlet fields is produced through dimension 5 decays of the charged components of $16_m$ close to the BBN epoch. These decays are interesting in their own right, as they may explain the observed Lithium abundances \cite{Jedamzik2008}.
On the other hand, dimension 6 decays between different components of the singlet supermultiplet may lead to observable astrophysical signals that we discuss  
in the rest of the paper. 

Note that these decays may go through operators generated by integrating out the heavy $U(1)_{B-L}$ gauge boson, or
by integrating out heavy $10_{\gut}$ fields. In the former case decays are flavor universal, while the latter generically lead to flavor non-universal decays.  In the case of flavor non-universal decays, since the decay rate scales as the fourth power of the coupling, it is easy to have decays to one flavor dominate over the rest.  Depending on the relative strength of gauge and superpotential couplings and the masses of the heavy fields, both possibilities can be realized.

One may  worry that this picture could be spoiled  by lower dimension operators, such as Kahler kinetic 
mixings  $10_{\gut}^{\dagger} 10_h$ and $16_m^{\dagger} 16_f$. However, these are forbidden by R-parity (under which $16_m$
is even, and $10_\gut$ is odd), and $m$-parity under which both $16_m$ and $10_\gut$ are odd.

For simplicity, in this paper we will focus on the case in which the scalar $\tilde s$ receives a TeV scale vev. 
In this case dimension 6 operators lead to two body decays of the singlet fields to the MSSM fields. We are thus lead to two interesting observations:
\begin{itemize}
\item  In this case dark matter decay products necessarily contain MSSM superpartners, because direct decays  of a scalar into two light fermions are
 suppressed by helicity. 
\item The production of superpartners, combined with the generic expectation that sleptons are lighter than squarks leads to decays dominantly into leptonic channels due to kinematics.
\end{itemize}
These lead to a possible connection between the branching fraction of dark matter and the spectrum of its decay products on the one hand, and the supersymmetric spectrum and the decay cascades of superpartners on the other.

\section{Astrophysical Signals}


\subsection{Electrons and Positrons}


GUT induced dark matter  decays lead to several generic expectations for electron/positron spectra. As discussed in the previous section, the dark matter is a combination of the scalar ($\tilde{s}$) and fermion ($s$) components of the $S_m$ superfield. The two body decay of dark matter will involve sleptons ($\tilde{l}$, the superpartner of a lepton) in the final state. The slepton then further decays to its partner lepton and the lightest supersymmetric particle (LSP), $\tilde{l} \to l + \text{LSP}$, leading to an injection spectrum of the lepton that is flat between a lower and upper edge~\cite{Arvanitaki:2008hq}. When dark matter is much heavier than the superpartners the lower edge is typically a few GeV and the upper edge is roughly at 
\begin{equation}
\label{edge}
E_{\mathrm{edge}}=\left(1-\frac{m^2_{\mathrm{LSP}}}{m^2_{\tilde l}}\right)\frac{m_{\mathrm {DM}}}{2}
\end{equation}
The sensitivity of the upper edge in the injection spectrum to the masses of superpartners may lead to interesting cross checks at the LHC.

Current data for the positron fraction from PAMELA~\cite{PAMELA} and the $e^-+e^+$ spectrum from the ATIC~\cite{ATIC} and HESS~\cite{HESS} experiments include interesting hints of a possible excess above background. Due to systematic uncertainties it is premature to interpret spectral shapes observed by ATIC and HESS, particularly above 100~GeV. However, this flux will be measured in the very near future by both Fermi and PAMELA with improved statistics and systematics. The HESS experiment will also extend its measurement to lower energies, overlapping with the excess observed by ATIC. 
It is thus interesting to consider several scenarios for the outcome of upcoming experiments and their implications in the context of our framework.

\begin{figure}
\begin{center}
\includegraphics[width=3 in]{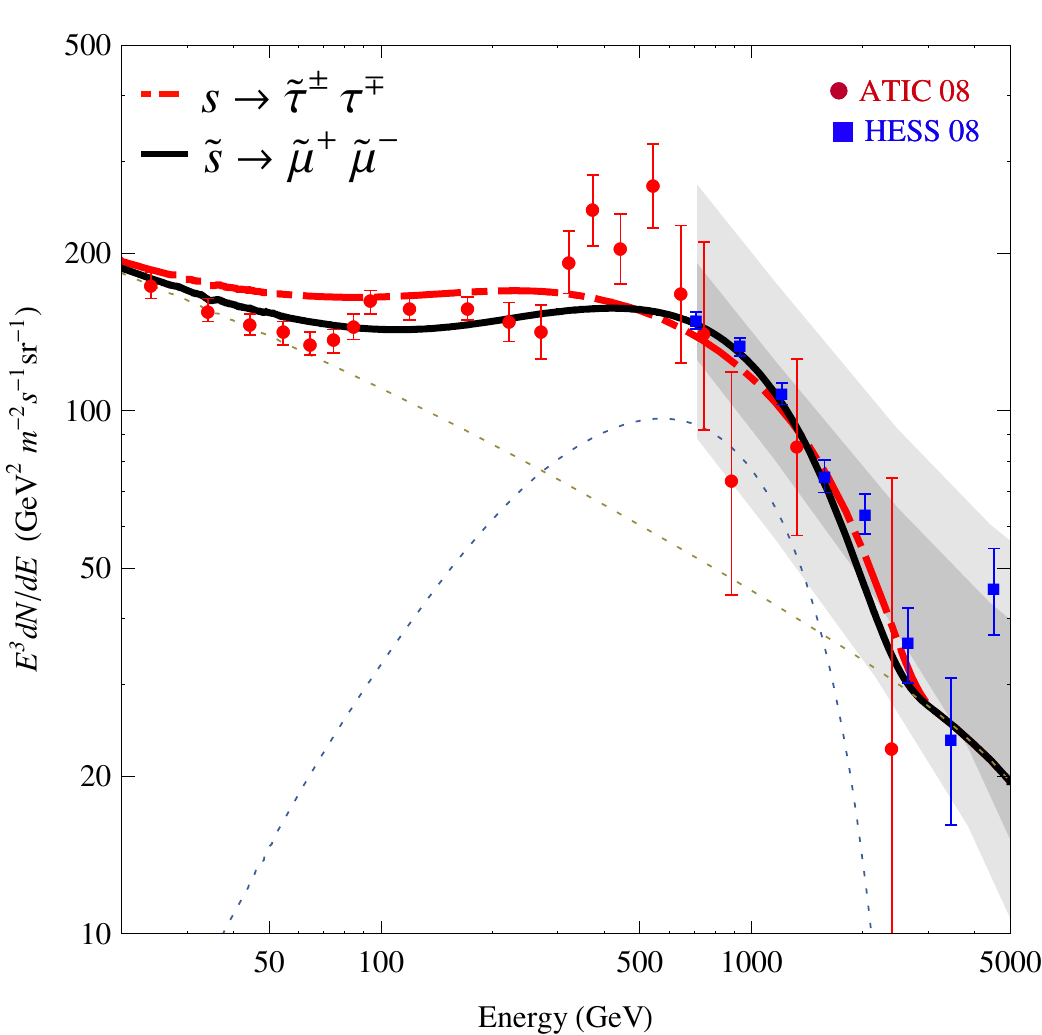}
\caption{ \label{fig-HESSian}
The electron+positron spectrum produced by the decay of dark matter ($s \text{ or } \tilde{s}$) to smuon pairs (solid black) and tau-stau pairs (dot-dashed red) as discussed in the text.
The HESS and ATIC data is shown by red squares and blus circles respectively. The systematic error of HESS is shown as a grey band and its superimposed energy uncertainty is shown as a lighter grey band. The background and smuon signal components of the flux are shown by dotted lines.}
\end{center}
\end{figure}

For concreteness we will focus on two scenarios of possible electron+positron spectra which are roughly consistent with current data, given the systematic uncertainties, but will be probed further soon: 

{\bf Scenario 1 - A smooth HESS signal} - The low energy data from ATIC  (below $\sim 75$ GeV) is used to estimate the astrophysical background. We then find decay channels to explain the higher energy HESS excess. Two simple examples that realize this  are shown in figure~\ref{fig-HESSian}.  
%

One possibility is that a heavy scalar (of order 6 TeV or more) is decaying to a pair of smuons which each further decay to a muon and a neutralino. The muons decay further to produce electrons and positrons.  
For example, a 8 TeV scalar decaying to 200 GeV smuons which then decay to 100 GeV neutralinos is shown in figure~\ref{fig-HESSian}. 
The spectral shape (see equation (\ref{edge})) produced by such a cascade fits the spectrum observed by HESS quite well.

A second possibility shown in figure~\ref{fig-HESSian} is that a heavy 6 TeV fermion decays to tau+stau, producing a similar spectral shape. A decay into the third family may be motivated by minimal flavor violation.  In this case the observed flux is dominated by the electrons and positrons produced in the subsequent decay of the hard ($\sim 3$ TeV) tau\footnote{It is generally interesting to point out that the HESS shape agrees well with dark matter decaying or annihilating to taus.}. The stau will produce an additional soft component to the spectrum which is subdominant. 

In this scenario, we take the lifetimes of the dark matter to be $10^{26}$ seconds for the smuon case and $ 6\times10^{25}$ seconds for the tau case. Though both of these possibilities produce a similar electron plus positron spectrum they may be distinguishable by gamma ray and neutrino observations, as discussed later in this paper. It is interesting to note that both of these possibilities require flavor non-universal decays of dark matter which may naturally be produced in the model discussed above. Furthermore, since the decay rate scales as the fourth power of the coupling, these flavor non-universal decays can be easily dominated by decays to one flavor over the others. 


{\bf Scenario 2 - Multiple features} - 
We now focus on some interesting spectra that consist of multiple features and that occur in simple scenarios of GUT induced dark matter decays. We will not necesarily assume that either the HESS or ATIC spectra are correct, but rather pick three examples to demonstrate some of the generic possibilities.

\begin{figure}
\begin{center}
\includegraphics[width=3 in]{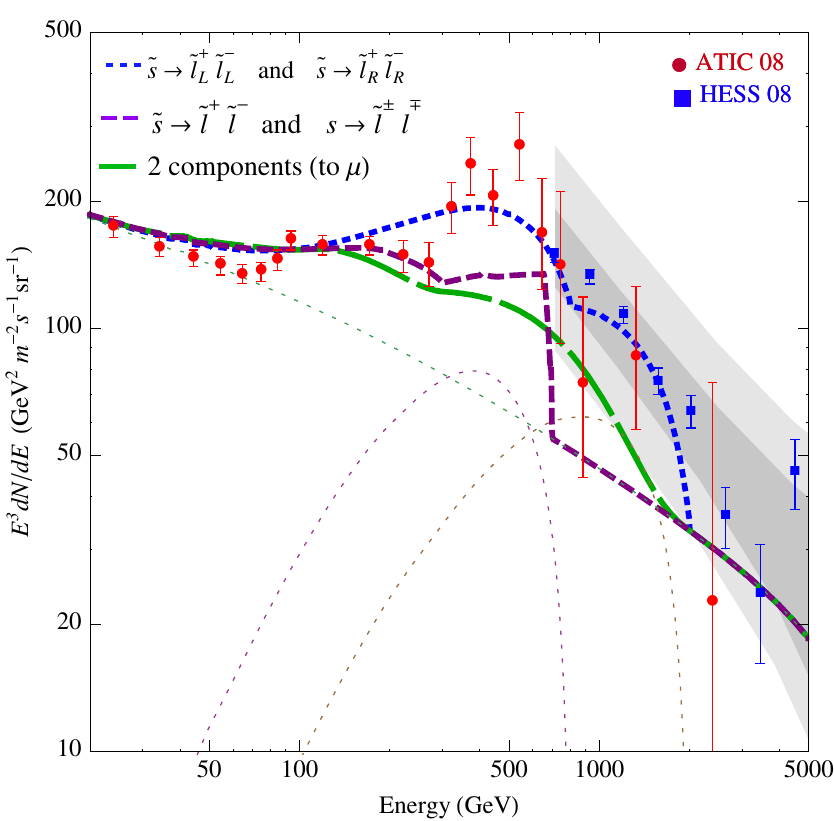}
\caption{ \label{fig-doubleaction}
The ``multi feature'' electron+positron spectra produced in simple dark matter decays scenarios. 
The short-dashed purple line is an example of multiple features arising from decays into leptons and sleptons. The blue dotted line is an example of a flavor universal decay to two sleptons of different masses. The long-dashed green line is for two DM components decaying to muons and smuons. 
The experimental data shown is similar to that in Figure 1. The two signal components and the background in the slepton case, as well as the background are shown as dotted lines.
}
\end{center}
\end{figure}

In the first example, dark matter decays  into leptons+sleptons via $s\to l^{\pm}\tilde l^{\mp}$ and to sleptons via $\tilde s\to \tilde l^{\pm} \tilde l^{\mp}$. The lepton final states lead to a hard high energy feature while the slepton final state leads to a smoother  one at lower energies. 
The purple short-dashed line of figure~\ref{fig-doubleaction} shows the flavor universal decay of a scalar and a fermion of mass 1.5 TeV and equal abundance. Here the slepton masses are  universal at 130 GeV and the LSP mass is 100 GeV. As shown in~\cite{Arvanitaki:2008hq}  a similar decay may yield a spectrum that is in remarkable agreement with the double feature observed by ATIC. 
Since the spectrum of superpartners and the mass of dark matter set the scale of both spectral features (see equation (\ref{edge})), a non-trivial cross check may be made once the LHC measures the masses of sleptons and the LSP.

Another possible way in which GUT induced dark matter decays can produce multiple spectral features is by various cascades of supersymmetric particles. 
For example, if dark matter decays to two sleptons of different mass ($\tilde l_L$ and $\tilde l_R$) both of which decay to a neutralino LSP, the upper edges of the injection spectra in the two cascades may be sufficiently different (see equation (\ref{edge})), leading to multiple features. In figure~\ref{fig-doubleaction} we show the spectrum produced by a flavor universal decay of dark matter into 145 GeV and and 230 GeV sleptons which subsequently decay to a 120 GeV LSP and a lepton (dotted blue line). The branching fraction into the heavier slepton is 20\%. Again, since both spectral features are set by relations of the form~{(\ref{edge}), this interpretation may be tested at the LHC.

Alternatively, multiple features may be produced by two cascades of the same slepton via different neutralino states.  For example, a similar spectrum (the dotted blue line in figure~\ref{fig-doubleaction}) may be produced by a 5 TeV scalar dark matter  decaying to sleptons with a mass of  200 GeV. The slepton has two dominant decay channels into two different neutralino states with masses of 170 and 100 GeV\footnote{It is unclear to what extent a hadronic decay of the next-to -lightest neutralino is in tension with the PAMELA anti-proton measurement~\cite{Grajek:2008pg}.
The next-to-lightest neutralino may also decay to taus without producing antiprotons if the lightest stau is accesible.}.

Spectral features can also arise from decays of two different dark matter particles with significantly different masses. Despite their different masses, the electron flux from the decays of these two particles can be comparable if their relic abundance is generated from the decays of another particle. For example, in the $SO(10)$ model discussed earlier, the relic number density $n_{\tilde{s}}$ and $n_s$ of the singlets $\tilde{s}$ and $s$ are generated through the decays of the components (with mass $m$) of the $16_m$ that are charged under the standard model. With singlet masses $m_{\tilde{s}}$ and $m_{s}$ for $\tilde{s}$ and $s$ respectively, their relic number densities satisfy  $\frac{n_{\tilde{s}}}{n_s} \sim \left(\frac{m - m_{\tilde{s}}}{m - m_{s}}\right)^3$. In this model, the $\tilde{s}$ and $s$ can decay only when $\tilde{s}$ develops a vev. Their respective dimension 6  decay rates $\Gamma_{\tilde{s}}$ and $\Gamma_s$ scale as  $\frac{\Gamma_{\tilde{s}}}{\Gamma_s} \sim \left(\frac{m_{\tilde{s}}}{m_{s}}\right)^3$. The relative electron flux from the two decays is  $\frac{n_{\tilde{s}} \Gamma_{\tilde{s}}}{n_s \Gamma_s} \sim \left(\frac{m_{\tilde{s}}}{m_{s}}\right)^3  \left(\frac{m - m_{\tilde{s}}}{m - m_{s}}\right)^3$. This ratio is $\OrderOne$ for $m_{\tilde{s}} + 0.4 \; m_{s} \lessapprox m  \lessapprox m_{\tilde{s}} + 2.2 \; m_{s}$ when $m_{\tilde{s}} \gg m_s$. For TeV scale  $m_{s}$ and $m_{\tilde{s}}$, this results in comparable electron fluxes  for  $m$ within a TeV of $m_{\tilde{s}} + m_s$. The observation of these features in the electron spectrum can thus lead to measurement of SUSY parameters and GUT physics.  In figure~\ref{fig-doubleaction}, we show the spectrum from the decays of a heavy 5 TeV scalar, $\tilde{s}$, and a 660 GeV fermion, $s$,  into $\tilde{s} \to \tilde{\mu}^{\pm}\tilde{\mu}^{\mp}$ and $s \to \mu^{\pm} \tilde{\mu}^{\mp}$.  The decay rates are $\Gamma_{\tilde{s}} = 8 \times 10^{-27} \text{ s}^{-1}$ and  $\Gamma_{s} = 3 \times 10^{-27} \text{ s}^{-1}$. The smuon was taken to be at 200 GeV and the LSP at 90 GeV. In this case, the heavier components of the $16_m$ are around $\sim 5.5$ TeV.

All of the scenarios described above are consistent with the qualitative shape of the PAMELA excess. The positron fraction for some of these cases is shown in figure~\ref{fig-PAMELA-all}.
\begin{figure}
\begin{center}
\includegraphics[width=3 in]{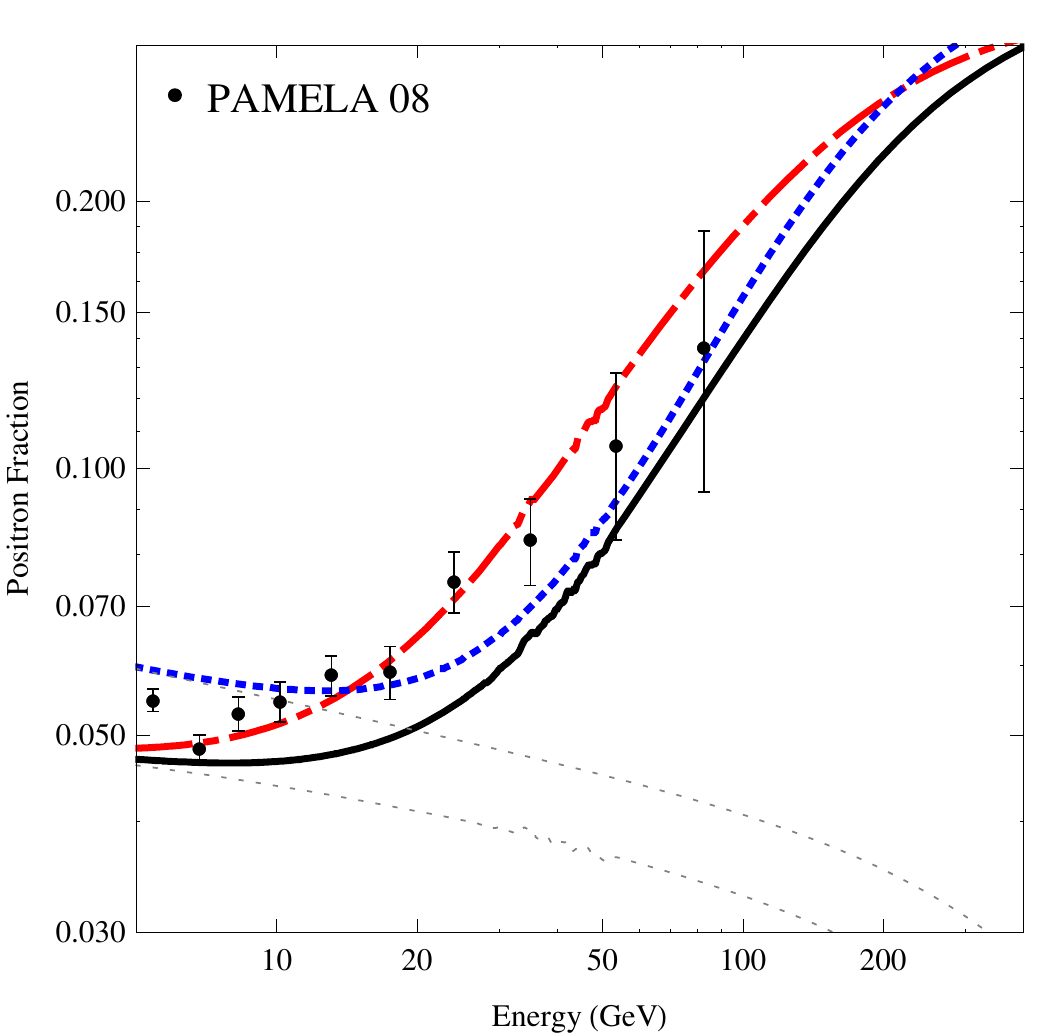}
\caption{ \label{fig-PAMELA-all}  
The positron fraction produced by some of scenarios discussed in the text - decays to smuons (Scenario 1, solid black), a universal lepton-slepton decay (Scenario 2, short-dashed purple) and decays to sleptons (Scenario 2, dotted black). The background positron fraction from two propagation models is shown in gray, thin dotted lines. The PAMELA data is shown (black circles).} 
\end{center}
\end{figure}
Before proceeding to photon signals we will discuss technical aspects of the figures above.
We used GALPROP~\cite{GALPROP} for generating backgrounds and for propagation of the dark matter signal. In figure~\ref{fig-HESSian}, we assumed the convective diffusion propagation model (DC) of~\cite{morselli}. In figure~\ref{fig-doubleaction},
a harder propagation model was used throughout (model B of~\cite{Cholis:2008hb}). The slope of the background electron flux, which may significantly affect both the positron fraction and the total flux was chosen to be $\sim -3.3$ at 20 GeV, in agreement with observations and the large uncertainties~\cite{DiracDM}. DarkSUSY~\cite{DarkSusy} was used for producing injection spectra.


\subsection{Diffuse Gamma-rays}
\label{Photons}

Any decay scenario that produces charged particles must produce gamma-rays from final state radiation (FSR) off those charged particles.  Fig. \ref{fig-FSRandICS} shows the gamma ray spectra from some of the models discussed above.  The shape of the FSR spectrum is directly related to the shape of the primary charged particle injection spectrum from dark matter decay \cite{Arvanitaki:2008hq}.  We do not include gamma-rays from inverse-compton scattering of starlight off the high energy electrons and positrons from the dark matter decay.  This does not usually give as hard a spectrum as FSR, and presumably falls off faster than FSR with galactic latitude as the density of starlight decreases off the plane of the galaxy (but see \cite{Moskalenko:1998gw}).  This contribution is difficult to calculate off the plane of the galaxy due to the anisotropic flux of starlight \cite{Igor, Moskalenko:1998gw}.  If a decay mode contains $\tau$'s we do include the full spectrum from these, e.g. the photons from $\pi^0$'s produced in the $\tau$ decay generated using \cite{DarkSusy}.

\begin{figure}
\begin{center}
\includegraphics[width=3.3 in]{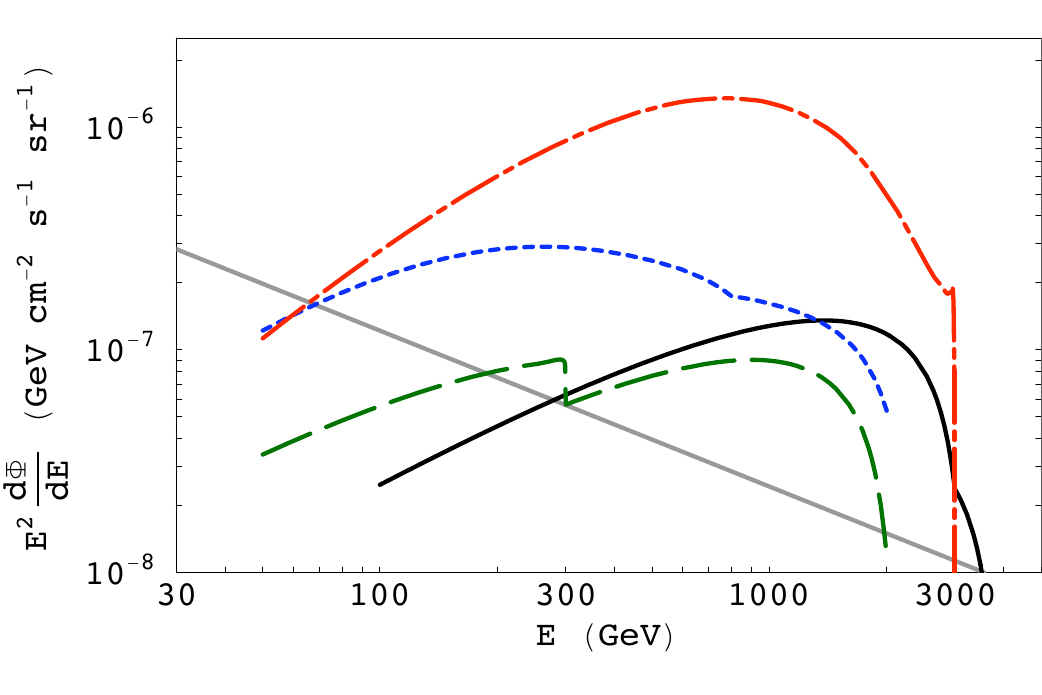}
\caption{ \label{fig-FSRandICS} The gamma-ray spectra from final state radiation and $\tau$ decay.  The solid (black) and dot-dashed (red) lines are as in Fig. \ref{fig-HESSian}.  The dashed (green) and dotted (blue) lines are as in Fig. \ref{fig-doubleaction}.  Only the signals are plotted, not signal plus background.  Gray is the expected background from \cite{Bergstrom:1997fj}.  These are shown at a galactic latitude $b = 60^\circ$ and longitude $l = 0^\circ$.}
\end{center}
\end{figure}

This diffuse gamma-ray signal can determine whether the observed electron/positron excesses do indeed arise from dark matter.  Dark matter decays give rise to an isotropic gamma-ray signal with a hard, high-energy spectral feature such as an edge coming from FSR.  The shape of the spectrum is the same everywhere across the sky.  The intensity varies slightly, with a dependence on the angle of observation that is determined by the known density of dark matter in the galactic halo and so is only uncertain at the galactic center.  If the electron/positron excesses arise from dark matter decay, the gamma-ray signal is probably strong enough to be observable at Fermi \cite{Arvanitaki:2008hq}.  Observation of such a signal would be impossible to explain by any known astrophysical mechanism other than dark matter.

These gamma-ray observations may even help distinguish decays from annihilations, since decays produce a more isotropic flux than that from annihilations.  Thus, the gamma-ray signal from decaying dark matter can be observed cleanly with measurements at high galactic latitude, off the plane of the galaxy, where there is little astrophysical background.

The gamma-ray spectrum carries important information about the nature of the DM decay that is obscured by the electron/positron spectra.  For example, the two scenarios in Fig.~\ref{fig-HESSian} produce almost identical electron/positron spectra even though the underlying high energy physics is different.  From Fig.~\ref{fig-FSRandICS} we see that these two are easily distinguished by their gamma-ray spectra.  Further, the shape of the FSR spectrum can provide a measurement of the mass of the dark matter particle and potentially the masses of its decay products such as superpartners.  For example, structure in the electron/positron spectrum in Fig.~\ref{fig-doubleaction} appears as well in the gamma-ray spectrum as in Fig.~\ref{fig-FSRandICS}.  In fact, the gamma-ray spectrum can provide extra information since FSR photons can come from any interior line in the decay chain that is on-shell.  For example, in the dashed (green) curves in Figs.~\ref{fig-doubleaction} and \ref{fig-FSRandICS} a 300 GeV muon is produced which is directly visible as an edge in the gamma-ray spectrum.  Thus, the gamma-ray spectrum can provide a probe of TeV scale physics complementary to the electron/positron spectra.

\subsection{Neutrinos}
When the Dark Matter particle decays to charged leptons, neutrinos will also be produced due to either $SU(2)$ invariance or subsequent decays of the produced muons or taus.  In Fig. \ref{neutrinoflux}, we present the $\nu_\mu$ flux in IceCube \cite{:2007td} for most of the above examples that predict neutrinos by SU(2) invariance. The solid black (dotted blue) line comes from $\tilde{s}\rightarrow  \tilde{\bar{\nu}}_{\mu (e)} + \tilde{\nu}_{\mu(e)}$, while the dot-dashed red line comes from $s\rightarrow \nu_\tau + \tilde{\nu}_\tau$. Because the neutrinos are produced at galactic distances, the flavor ratios on the earth are 1:1:1. We do not include the flux produced by charged lepton decays as they are subdominant. We take a bin size of 0.2 in $\text{Log}_{10}\left(\frac{E}{\text{GeV}}\right)$ as it appears in \cite{:2007td}, but the energy resolution could be as low as 0.4 in $\text{Log}_{10}\left(\frac{E}{\text{GeV}}\right)$ \cite{subir}. In the 100 GeV to several TeV range there is a large atmospheric neutrino background that drops rapidly with energy as a power law, $E^{-3}$. For example, taking into account the effective area of IceCube \cite{subir} and integrating over a bin of size 0.4 in $\text{Log}_{10}\left(\frac{E}{\text{GeV}}\right)$ centered around the neutrino energy, $5\sigma$ discovery of a one TeV neutrino line could be possible within roughly a year of observation time, when the Dark Matter lifetime is $10^{26}$ sec. In this estimate we have not included systematics in the measurement process. Spectral information or distinguishing between different scenarios is going to be harder to deduce because of the atmospheric neutrino background and IceCube's energy resolution. It is worth noting though that both the IceCube energy resolution as well as the atmospheric background subtraction could be greatly improved in the near future, increasing the potential of the experiment to observe and study astrophysical signals of TeV scale Dark Matter.


\begin{figure}
\begin{center}
\includegraphics[width=3.3 in]{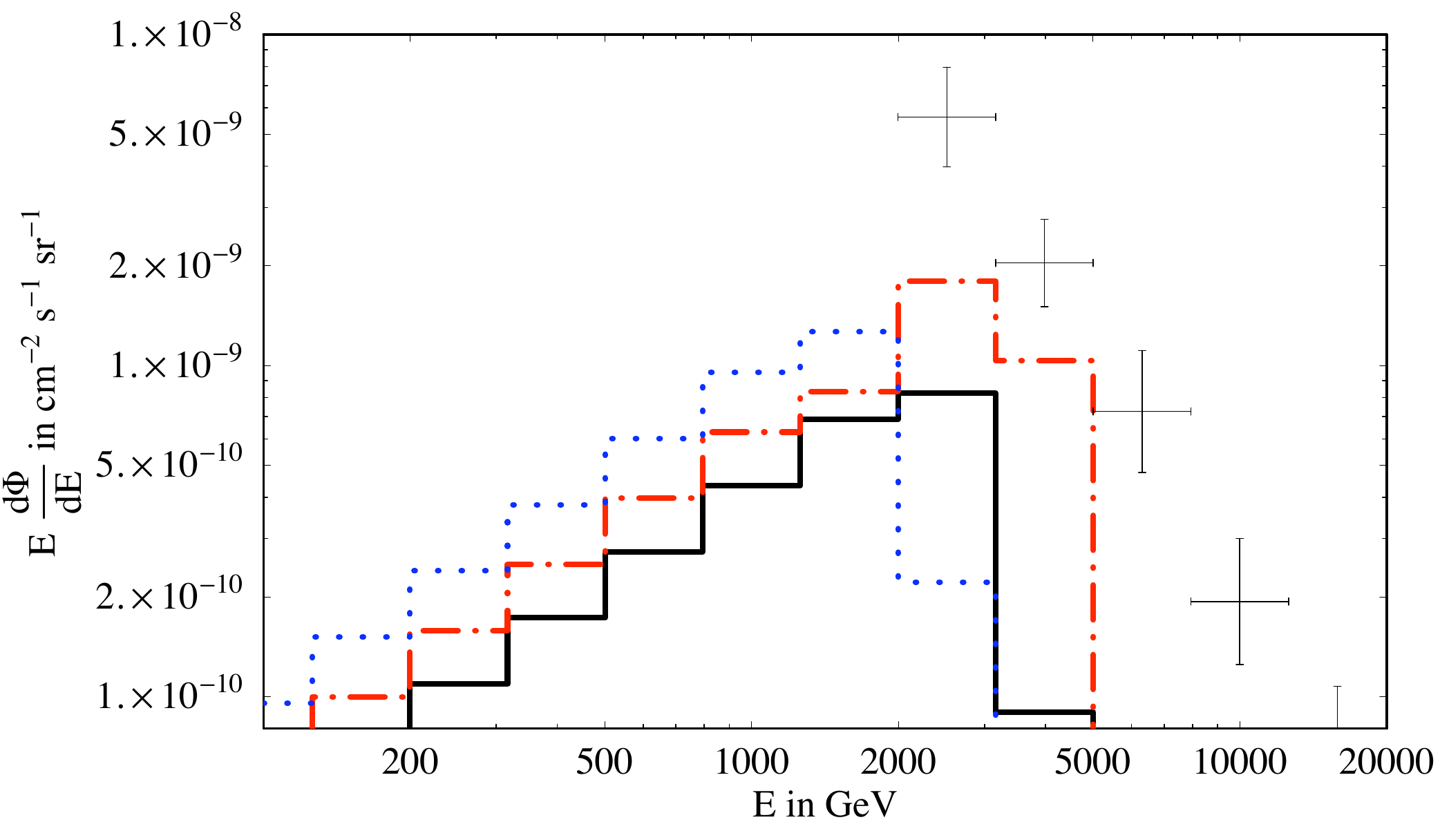}
\caption[neutrinos]{The average $\nu_\mu$ flux produced from a Dark Matter particle decay in different examples. The solid (black) and dot-dashed (red) lines are as in Fig. \ref{fig-HESSian}.  The dotted (blue) line is as in Fig. \ref{fig-doubleaction}. AMANDA data is shown in gray \cite{:2007td}. The bin size is taken to be 0.2 in $\text{Log}_{10}\left(\frac{E}{\text{GeV}}\right)$, as it appears in \cite{:2007td}.}
\label{neutrinoflux}
\end{center}
\end{figure}

\section{Conclusions}
The decaying dark matter scenarios explored in this paper can naturally explain the cosmic ray spectra observed at HESS, PAMELA and ATIC without the need for astrophysical boost factors or additional new energy scales \cite{Arvanitaki:2008hq, Nardi:2008ix, MaximPospelov, YanagidaOne, YanagidaTwo, YanagidaThree, Chen:2009ew, Frampton:2009yc, Ishiwata:2009vx, Shirai:2009kh, Goh:2009wg, Takahashi:2009mb, Foadi:2008qv, Hamaguchi:2008ta, MinimalDarkMatter, NimaDougTracyWeiner, Kadastik:2009dj, Barger:2009yt, Yin:2008bs, Bae:2009bz, Kyae:2009jt, Covi:2008jy, Ibarra:2008qg, Ibarra:2007wg, Berezinsky:1996pb}.  In particular, the softer electron spectrum observed by HESS can fit naturally with the observations of PAMELA.  Substructures in the electron spectrum, correlated with substructures in the photon spectrum, are also possible in these theories. Such substructure may also be correlated with the spectrum of superpartners observed at the LHC. Observation of these substructures could open a window into the spectrum of TeV mass particles.  The HESS observation suggests a heavy ($\gtrsim$ TeV) DM mass, too heavy to be produced at the LHC.  In this case, astrophysical observations would provide our only probe of dark matter at such a high energy scale.

The same theories typically also contain particles decaying during big bang nucleosynthesis through dimension 5 operators with lifetime $\tau \sim 8 \pi \frac{\Mgut^2}{m^3} \approx 7 ~\s .$  Such decays are recorded by a change in the primordial light element abundances and may explain the anomalous observed Li abundances, opening another window to unification \cite{Arvanitaki:2008hq}.


\section*{Acknowledgments}
We would like to thank  Bill Atwood, Patrick Fox,  Giorgio Gratta, Francis Halzen, Dan Hooper, Graham Kribs,  John March-Russell, Peter Michelson, Igor Moskalenko, Hitoshi Murayama, Roger Romani,  Subir Sarkar, Bob Wagoner and Neal Weiner for useful discussions.


\begin{thebibliography}{10}
\expandafter\ifx\csname url\endcsname\relax
  \def\url#1{{\tt #1}}\fi
\expandafter\ifx\csname urlprefix\endcsname\relax\def\urlprefix{URL }\fi


\bibitem{Arvanitaki:2008hq}
  A.~Arvanitaki, S.~Dimopoulos, S.~Dubovsky, P.~W.~Graham, R.~Harnik and S.~Rajendran,
  arXiv:0812.2075 [hep-ph].
  
  
\bibitem{HESS}
  H.~E.~S.~S.~Collaboration,
  arXiv:0811.3894 [astro-ph].
  
\bibitem{Jedamzik2008}
  S.~Bailly, K.~Jedamzik and G.~Moultaka,
  arXiv:0812.0788 [hep-ph].

\bibitem{PAMELA}
  O.~Adriani {\it et al.}  [PAMELA Collaboration],
  arXiv:0810.4995 [astro-ph].

\bibitem{ATIC}
  J.~Chang {\it et al.},
  Nature {\bf 456}, 362 (2008).
  
  




\bibitem{GALPROP}
  http://galprop.stanford.edu/web\_galprop/\\galprop\_home.html;
  A.~W.~Strong and I.~V.~Moskalenko,
  arXiv:astro-ph/9906228;
  A.~W.~Strong and I.~V.~Moskalenko,
  arXiv:astro-ph/0106504.



\bibitem{morselli}
  A.~M.~Lionetto, A.~Morselli and V.~Zdravkovic,
  JCAP {\bf 0509}, 010 (2005)
  [arXiv:astro-ph/0502406].
  
\bibitem{Cholis:2008hb}
  I.~Cholis, L.~Goodenough, D.~Hooper, M.~Simet and N.~Weiner,
  arXiv:0809.1683 [hep-ph].


\bibitem{DiracDM}
  R.~Harnik and G.~D.~Kribs,
  arXiv:0810.5557 [hep-ph].

\bibitem{DarkSusy}
  P.~Gondolo, J.~Edsjo, P.~Ullio, L.~Bergstrom, M.~Schelke and E.~A.~Baltz,
  JCAP {\bf 0407}, 008 (2004)
  [arXiv:astro-ph/0406204].

\bibitem{Grajek:2008pg}
  P.~Grajek, G.~Kane, D.~Phalen, A.~Pierce and S.~Watson,
  arXiv:0812.4555 [hep-ph].


\bibitem{Bergstrom:1997fj}
  L.~Bergstrom, P.~Ullio and J.~H.~Buckley,
  Astropart.\ Phys.\  {\bf 9}, 137 (1998)
  [arXiv:astro-ph/9712318].
  
\bibitem{Moskalenko:1998gw}
  I.~V.~Moskalenko and A.~W.~Strong,
  Astrophys.\ J.\  {\bf 528}, 357 (2000)
  [arXiv:astro-ph/9811284].
  
\bibitem{Igor}
I. Moskalenko, private communication.



\bibitem{:2007td}
    [IceCube Collaboration],
  arXiv:0711.0353 [astro-ph].

\bibitem{subir}
S. Sarkar, private communication.


\bibitem{Nardi:2008ix}
  E.~Nardi, F.~Sannino and A.~Strumia,
  arXiv:0811.4153 [hep-ph].

\bibitem{MaximPospelov}
M.~Pospelov and M.~Trott,
[arXiv:0812.0432]

\bibitem{YanagidaOne}
C.~C.~Chen, F.~Takahashi and T.~T.~Yanagida,
[arXiv:0809.0792]

\bibitem{YanagidaTwo}
C.~C.~Chen, F.~Takahashi and T.~T.~Yanagida,
[arXiv:0811.0477]

\bibitem{YanagidaThree}
K.~Hamaguchi, E.~Nakamura, S.~Shirai and T.~T.~Yanagida, 
[arXiv:0811.3357]

\bibitem{Chen:2009ew}
  S.~L.~Chen, R.~N.~Mohapatra, S.~Nussinov and Y.~Zhang,
  arXiv:0903.2562 [hep-ph].

\bibitem{Frampton:2009yc}
  P.~H.~Frampton and P.~Q.~Hung,
  arXiv:0903.0358 [hep-ph].


\bibitem{Ishiwata:2009vx}
  K.~Ishiwata, S.~Matsumoto and T.~Moroi,
  arXiv:0903.0242 [hep-ph].


\bibitem{Shirai:2009kh}
  S.~Shirai, F.~Takahashi and T.~T.~Yanagida,
  arXiv:0902.4770 [hep-ph].


\bibitem{Goh:2009wg}
  H.~S.~Goh, L.~J.~Hall and P.~Kumar,
  arXiv:0902.0814 [hep-ph].


\bibitem{Takahashi:2009mb}
  F.~Takahashi and E.~Komatsu,
  arXiv:0901.1915 [astro-ph].


\bibitem{Foadi:2008qv}
  R.~Foadi, M.~T.~Frandsen and F.~Sannino,
  arXiv:0812.3406 [hep-ph].


\bibitem{Hamaguchi:2008ta}
  K.~Hamaguchi, S.~Shirai and T.~T.~Yanagida,
  Phys.\ Lett.\  B {\bf 673}, 247 (2009)
  [arXiv:0812.2374 [hep-ph]].
  

\bibitem{MinimalDarkMatter}
  M.~Cirelli and A.~Strumia,
  arXiv:0808.3867 [astro-ph].
\bibitem{NimaDougTracyWeiner}
N.~Arkani-Hamed, D.~P.~Finkbeiner, T.~R.~Slatyer and N.~Weiner,
[arXiv:0810.0713]

\bibitem{Kadastik:2009dj}
  M.~Kadastik, K.~Kannike and M.~Raidal,
  arXiv:0903.2475 [hep-ph].
 
\bibitem{Barger:2009yt}
  V.~Barger, Y.~Gao, W.~Y.~Keung, D.~Marfatia and G.~Shaughnessy,
  arXiv:0904.2001 [hep-ph].
   
\bibitem{Yin:2008bs}
  P.~f.~Yin, Q.~Yuan, J.~Liu, J.~Zhang, X.~j.~Bi and S.~h.~Zhu,
  Phys.\ Rev.\  D {\bf 79}, 023512 (2009)
  [arXiv:0811.0176 [hep-ph]].

\bibitem{Bae:2009bz}
  K.~J.~Bae and B.~Kyae,
  arXiv:0902.3578 [hep-ph].
 
\bibitem{Kyae:2009jt}
  B.~Kyae,
  arXiv:0902.0071 [hep-ph].
   
\bibitem{Covi:2008jy}
  L.~Covi, M.~Grefe, A.~Ibarra and D.~Tran,
  JCAP {\bf 0901}, 029 (2009)
  [arXiv:0809.5030 [hep-ph]].
 
\bibitem{Ibarra:2008qg}
  A.~Ibarra and D.~Tran,
  JCAP {\bf 0807}, 002 (2008)
  [arXiv:0804.4596 [astro-ph]].
  
  
  
\bibitem{Ibarra:2007wg}
  A.~Ibarra and D.~Tran,
  Phys.\ Rev.\ Lett.\  {\bf 100}, 061301 (2008)
  [arXiv:0709.4593 [astro-ph]].
  
   
  
\bibitem{Berezinsky:1996pb}
  V.~Berezinsky, A.~S.~Joshipura and J.~W.~F.~Valle,
  Phys.\ Rev.\  D {\bf 57}, 147 (1998)
  [arXiv:hep-ph/9608307].


\end{thebibliography}
\end{document}